\documentclass[12pt]{article}
\usepackage{epsfig}
\usepackage{amssymb}
\begin{document}

\vspace*{20mm}
\begin{flushright}
CERN-TH/2000.015
\end{flushright}
\vspace*{1cm}
\begin{center}
{\bf CONTINUATION OF DIRECT PRODUCTS OF DISTRIBUTIONS}
\vspace*{1cm}

{\bf A. Peterman }\\
\vspace{0.3cm}
Theoretical Physics Division,
CERN \\
CH -- 1211 Geneva 23\\
\vspace*{2cm}
 \end{center}

\setcounter{page}{1}
\pagestyle{plain}

\begin{center}{\bf Preamble}\end{center}
If, in some problems, one has to deal with the ``product'' of
distributions
$\rm f_i$ (also called
generalized functions) $\rm\overline T = \Pi^m_{i=1} f_i$, this product
has
a priori no definite meaning
as a functional $(\rm \overline T, \varphi) $ for $\rm\varphi \in S$. But
if $\rm x^{\kappa +1}
\Pi^m_{i=1} f_i$ exists, whatever the associativity is  between some
powers
$\rm r_i$ of $\rm x$ ($\rm
r_i
\in \Bbb N, \sum_i r_i\leq \kappa  +1, r_i \geq 0$) and the various $\rm
f_i$, then a continuation
of the  linear functional $\rm \overline T$ from $\rm M$  onto $\rm
S^{(N)}$  for some $\rm N$
is shown to exist\footnote{$\rm M$ is a closed subspace of $\rm S^N$ for
some $\rm N$. It is a Banach
space with norm
$\rm \|
\|_N $.} in such a way that $\rm x^{\kappa +1} \overline T$ is defined
unambiguously, and $\rm (\overline
T,
\varphi), \varphi \in S$, significant, though not unique.
\vspace*{4cm}
\begin{flushleft}
CERN-TH/2000-015\\
January 2000
\end{flushleft}
\vfill\eject

\section{Existence}

In the sense of convergence in the space $\rm S^*$ (distributions),
$$\rm f_\kappa = \lim_{y \rightarrow 0,\; y \in C^+} F^y_\kappa (x)
\;;\;\;
\kappa = 1,2, \cdots ,
m\;,$$
 with $\rm F^y_\kappa (x) = f^+_\kappa (x +iy) - f^-_\kappa (x - i y)$,
$\bigg(\rm f^\pm_\kappa (x)$ are
holomorphic in tabular domains $\rm T^{C^\pm_R}$ and satisfy
\begin{equation}
|f (x + i y) | \leq C (R^\prime, C^\prime) |y|^{-\alpha} (1+|x|)^\beta
\end{equation}
and
$$
z \in R^n + i (C^\prime
\cap U (0, R^\prime))
$$
$\rm \alpha , \beta \geq 0$, independent of $\rm R^\prime$ and $\rm
C^\prime$. From this, it follows that
there exists in $\rm S^*$ a unique boundary value
$$\left.{\rm f(x) = \lim_{y \rightarrow 0 , \; y \rightarrow c} f(x + iy)
 \in S^{(m) *}; \; m= \alpha +\beta +
n +3\;.}\right)$$

Let us suppose that for arbitrary  $\rm \varphi \in S$ there exists a
finite limit
\begin{equation}
\lim_{y \rightarrow 0, \;y\in C^+} \int  F_1^y (x) \cdots F^y_m (x) \cdot
\varphi (x) dx
\end{equation}
independent of the sequence $\rm y \rightarrow 0 , \; y \in C^+$. Then,
since the space $\rm S^*$ is
dense, this limit defines a distribution in $\rm S^*$ which we call the
product $\rm f_1 . f_2 .\cdots
.f_m$ of the distributions $\rm f_1, f_2,\cdots , f_m$. Thus
\begin{equation}
f_1. f_2. \cdots . f_m = \lim_{y\rightarrow 0 , \; y \in C^+} F^y_1 \cdots
F^y_m\;\;({\rm in}\; S^*)
\end{equation}
if the limit of the RHS exists and is independent of the sequence $\rm y
\rightarrow 0 , y \in C^+$.
This product is obviously commutative and associative. So  the set of
boundary values that are
holomorphic in $\rm T^{C^+_R}$ and satisfy (1) constitute a commutative
ring with unity, without zero
divisors with respect to the multiplication defined above.

We note that the existence of the lim in (2) for $\rm \varphi \in S$
implies the existence of the
limit in (3) with respect to the norm of the functional in $\rm S^{(N)*}$
for some $\rm N$, which depends
on
$\rm f_1 \dots f_m$ (notice that weak convergence in $\rm S^*$ implies
strong convergence).

\section{General case}

Suppose now that (2) does not exist for all $\rm \varphi \in S$, but that
it exists for all $\rm \varphi$
in a closed subspace  $\rm M$ of $\rm S^{(N)}$ for some $\rm N$. (Since
$\rm M$ is closed in $\rm
S^{(N)}$ it is a Banach space with norm $\rm \|\;\|_N$). From the
Banach-Steinhaus theorem, (2) defines a
continuous linear functional $\rm \overline T$ on $\rm M$. We use now the
term `product' $\rm f_1.
\cdots . f_m$ of the distributions $\rm f_1, f_2, \cdots , f_m$ for
\underline{any} continuous linear
functional in the space $\rm S^{(N)*}\subset S^*$ that is a
\underline{continuation} of $\rm \overline T$
from
$\rm M$ to
$\rm S^{(N)}$. According to the Hahn-Banach theorem, such an extension
always exists but is not
\underline{unique} in general.

We shall concentrate now on the case of those $\varphi$ in $\rm S^{(N)}$
that vanish together with
all derivatives of order $\rm p\leq N$ inclusively, at $\rm x=0$. In this
case, \underline{all}
continuations
  $\rm f_1. f_2. \cdots . f_m$ of $\rm \overline T$ from $\rm M$ onto $\rm
S^{(N)}$ are given by
\begin{equation}
(f_1. f_2 . \cdots . f_m, \varphi ) = (\overline{T}, \overline{\varphi}) +
\sum_{\kappa \leq p}
c_\kappa (\delta^{(\kappa)}, \varphi)
\end{equation}
where
$$\rm \overline{\varphi} (x) = \varphi (x) - \sum_{\kappa \leq p}
\varphi^{(\kappa )} (o) \omega (x)
\frac{x^\kappa}{\kappa^!}$$
and $\rm \omega (x)$ is an arbitrary function, $\rm \omega \in S$,
identically equal to 1 in a
neighbourhood of the point $\rm x =0$; the $\rm c_\kappa$ are arbitrary
constants. (Notice that the
extension (4) is actually independent of $\rm \omega (x)$).

In conclusion, the formula (4) represents the desired result, given at the
end of the preamble
with $\rm \sum_{\kappa \leq p} c_\kappa \delta^{(\kappa)}$ the general
solution of $\rm (f_1. \cdots .
f_m, \varphi) =0$ and $\rm ( {T}, \overline{\varphi} ) = (\overline{T},
x^{\kappa +1} \psi) =
(x^{\kappa +1} \overline{T}, \psi ), \psi \in S$, a particular solution of
$\rm (f_1.\cdots . f_m,
\varphi )$.

It is therefore shown that the solution (4) is not unique, the $\rm
c_\kappa$ being arbitrary
constants.
\end{document}